\documentclass[12pt,a4paper]{article}

\usepackage{amsmath}
\usepackage{amsthm,amssymb}

\usepackage{graphicx} 

\newtheorem{theorem}{Theorem}
\newtheorem{proposition}[theorem]{Proposition}
\newtheorem{definition}[theorem]{Definition}

\begin{document}
\title{Scaling of the Atmosphere of Self-Avoiding Walks}
\author{A L Owczarek$^1$ and T Prellberg$^2$\\
  \footnotesize
  \begin{minipage}{13cm}
    $^1$ Department of Mathematics and Statistics,\\
    The University of Melbourne, Victoria~3010, Australia.\\
    \texttt{a.owczarek@ms.unimelb.edu.au}\\[1ex] 
$^2$ School of Mathematical Sciences\\
Queen Mary, University of London\\
Mile End Road, London E1 4NS, UK\\
\texttt{t.prellberg@qmul.ac.uk}
\end{minipage}
}

\maketitle  

\begin{abstract}
The number of free sites next to the end of a self-avoiding walk is known as the \emph{atmosphere}. 
The average atmosphere can be related to the number of configurations. Here we study the distribution 
of atmospheres as a function of length and how the number of walks of fixed atmosphere scale. 
Certain bounds on these numbers can be proved. We use Monte Carlo estimates to verify our conjectures. Of particular interest 
are walks that have zero atmosphere, which are known as \emph{trapped}. We demonstrate that these 
walks scale in the \emph{same} way as the full set of self-avoiding walks, barring an overall constant factor.
 
\end{abstract}

\newpage

Consider an $n$-step self-avoiding walk $\omega=(\omega_0,\omega_1,\ldots,\omega_n)$ 
with $n+1$ sites $\omega_i\in\mathbb Z^d$ for $d\geq2$, and steps having unit length,
i.e. $|\omega_{i+1}-\omega_i|=1$.

\begin{figure}[ht!]
\label{figure1}
\centering
\includegraphics[width=0.4\textwidth]{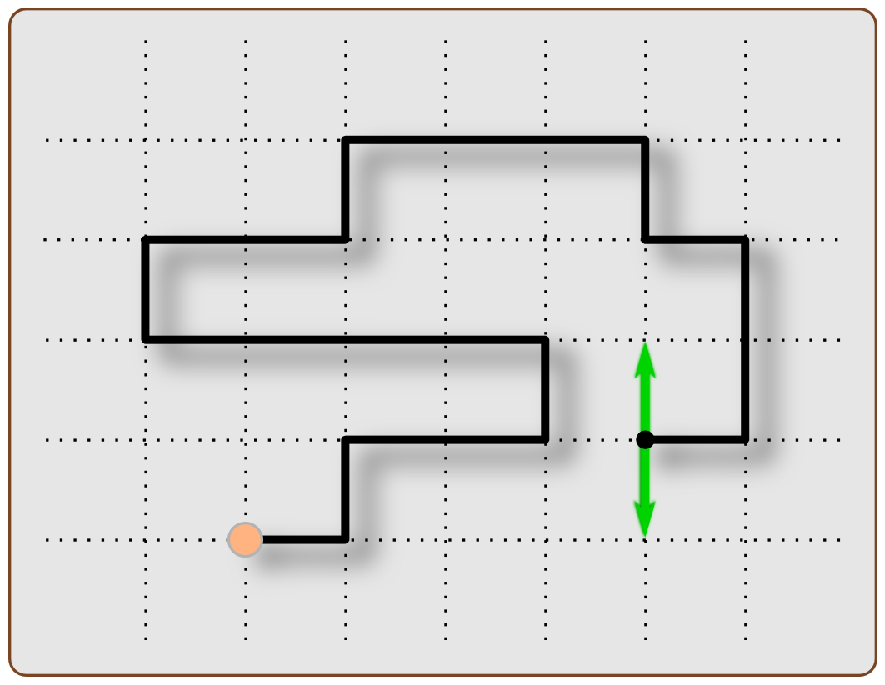}
\caption{A self-avoiding walk on the square lattice $\mathbb Z^2$ with
$n=20$ steps and an atmosphere $a=2$.}
\end{figure}

The number of edges that can be appended to the last visited vertex $\omega_n$ to 
create an $(n+1)$-step is called the \emph{atmosphere} of the walk $\omega$. 
Clearly the smallest value of the atmosphere is zero, in which case the walk is 
called \emph{trapped}. A zero-step self-avoiding walk has atmosphere $2d$, and for
$n\geq1$ any $n$-step self-avoiding walk has atmosphere of at most $2d-1$.

We partition the set of $n$-step self-avoiding walks by the value of their atmosphere.
Denote by $c_n$ the number
of $n$-step self-avoiding walks starting at $\omega_0=0$, and by $c_n^{(a)}$ the
number of $n$-step self-avoiding walks starting at $\omega_0=0$ with atmosphere $a$.

The subject of this paper is the fraction of $n$-step self-avoiding walks with fixed
atmosphere, 
\begin{equation}
p_n^{(a)}=\frac{c_n^{(a)}}{c_n}\;,
\end{equation}
 and its limiting behaviour as $n\to\infty$.

The atmosphere was introduced by Rechnitzer and Janse van Rensburg \cite{rechnitzer2002a} for self-avoiding walks. Since it is well established theoretically, though not proven, that
\begin{equation}
c_n \sim A_1 \mu^n n^{\gamma-1}
\label{cn-scale}
\end{equation}
they immediately pointed out that the \emph{mean} atmosphere could be used to estimate the connective constant $\mu$ and exponent $\gamma$ since 
\begin{equation}
\langle a \rangle = \sum_a a p_n^{(a)} =  \frac{\sum_a a c_n^{(a)}}{c_n} = \frac{c_{n+1}}{c_n}\;.
\end{equation}
Subsequently they have generalised this elegant idea to self-avoiding walks near walls \cite{rechnitzer2004a} and recently to self-avoiding polygons \cite{janse2008a}.

Motivating this paper is recent interest in various subsets of  walks that do not \emph{trap} \cite{guttmann2007,bousquet2008} including prudent walks \cite{turban1987,santra2001}. The question that naturally arises for each class of walks is \emph{``Does  the numbers of walks in the class scale in the same way as full set of self-avoiding walks"}; that is, as in equation~(\ref{cn-scale}) with the same value of $\mu$ and the same value of $\gamma$.

It is then natural to consider the limits
\begin{equation}
p^{(a)} = \lim_{n\to\infty} \frac{c_n^{(a)}}{c_n}\;.
\label{limits}
\end{equation}
In this paper we demonstate results in the literature \cite{madras1993,madras1988} can be used to prove
\begin{equation}
\liminf_{n\to\infty}\frac{c_n^{(a)}}{c_n}>0\;,
\label{liminfs}
\end{equation}
and so that if one assumes the limit  exists then one has immediately
 that
 \begin{equation}
 p^{(a)} >0\;.
 \end{equation}
 Going further and assuming the scaling in  equation~(\ref{cn-scale})  we can therefore predict
 \begin{equation}
c_n^{(a)} \sim A_1^{(a)} \mu^n n^{\gamma-1}\;,
\label{cna-scale}
\end{equation}
where
 \begin{equation}
 A_1^{(a)} =   p^{(a)} A_1\;.
 \end{equation}
  Hence walks that trap, that is, those with atmosphere zero will have the same connective constant and same value of exponent $\gamma$ as all self-avoiding walks. Conversely the same is true for walks that do not trap.

Furthermore we use the Monte Carlo Algorithm flatPERM\cite{prellberg2004}  to provided estimates for $p_n^{(a)}$ up to length $n=512$ and use the corrections-to-scaling to extrapolate high-precision estimates of $ p^{(a)} $.

\paragraph{Theory}
Intuitively one would expect that the probability that the end of
a long self-avoiding walk $\omega$ ends at the corner of a cube $Q$ 
of fixed size, but otherwise does not intersect with the cube, is non-zero.
Therefore one can append to this walk a short walk contained within the cube 
with fixed atmosphere. Using a related argument, we shall prove below  that
the \emph{limes inferior} of $p_n^{(a)}$ is bounded away from zero. 

Our main result is the following theorem.

\begin{theorem}
\label{theorem1}
Let $0\leq a<2d$. Then 
$$\liminf_{n\to\infty}p_n^{(a)}>0\;,$$
and
$$\limsup_{n\to\infty}p_n^{(a)}<1\;.$$
\end{theorem}

This theorem is very similar to a result concerning tail patterns of self-avoiding walks.

\begin{definition}
A self-avoiding walk $P=(p_0,p_1,\ldots,p_k)$ is called a \emph{tail pattern} if there is a self-avoiding walk
$\omega=(\omega_0,\omega_1,\ldots,\omega_n)$ such that $P=(\omega_{n-k},\omega_{n-k+1},\ldots,\omega_n)$.
$P$ is called a \emph{proper tail pattern} if for all sufficiently large $n$ there is a self-avoiding walk having
$P$ as a tail pattern.
\end{definition}

Let $c_n[P]$ be the number of $n$-step self-avoiding walks with tail pattern $P$.
We have the following result, which is taken from \cite{madras1993}.

\begin{proposition}
\label{madras}
If $P$ is a proper tail pattern then
$$\liminf_{n\to\infty}\frac{c_n[P]}{c_n}>0\;.$$
\end{proposition}

When applying this proposition for the proof of Theorem \ref{theorem1}, we find that the main difficulty 
is that the atmosphere of a proper tail pattern $P$ and the atmosphere of a self-avoiding walk having $P$
as a tail pattern can be different. 

\begin{proof}[Proof of Theorem \ref{theorem1}]
It is sufficient to show that for any atmosphere $a$ with $0\leq a<2d$ there exist a proper
tail pattern $P_a$ with atmosphere $a$ and the additional property that any sufficiently long 
self-avoiding walk having $P_a$
as a tail pattern has also atmosphere $a$. The first inequality then follows as an immediate 
consequence of Proposition \ref{madras}, as $c_n^{(a)}\geq c_n[P_a]$, and therefore
$$\liminf_{n\to\infty}p_n^{(a)}=\liminf_{n\to\infty}\frac{c_n^{(a)}}{c_n}\geq\liminf_{n\to\infty}\frac{c_n[P_a]}{c_n}>0\;.$$
The upper bound follows now from the observation that $\sum_{a=0}^{2d-1}p_n^{(a)}=1$, as
\begin{align*}
\limsup_{n\to\infty}p_n^{(a)}=&\limsup_{n\to\infty}\left(1-\sum_{a'\neq a}p_n^{(a')}\right)\\
\leq&1+\sum_{a'\neq a}\limsup_{n\to\infty}\left(-p_n^{(a')}\right)\\
=&1-\sum_{a'\neq a}\liminf_{n\to\infty}p_n^{(a')}<1\;.
\end{align*}
The pattern $P_a$ is constructed as follows. Consider a self-avoiding walk $\omega$ which starts at the origin, 
visits precisely $2d-a$ neighbouring vertices, and ends on the corner of a cube $Q$ containing it.
Continue $\omega$ by a self-avoiding walk $\omega'$ on the set of vertices $\partial Q$ which have distance 1 from $Q$,
and which is Hamiltonian on $\partial Q$. The pattern $P_a$ which is obtained by reversal of steps is a proper
tail pattern. It has atmosphere $a$, and by construction the atmosphere of any sufficiently long
self-avoiding walk having $P$ as a tail pattern must also be $a$.
\end{proof}

Note that the idea of this proof generalises easily to different notions of atmosphere, such as a $k$-step atmosphere
given by the number of $(n+k)$-step self-avoiding walks which can be grown from an $n$-step self-avoiding walk.

\paragraph{Simulation results}
Backed up by the rigorous results, we now consider results of simulations for
self-avoiding walks on the square lattice $\mathbb Z^2$. Using flatPERM, 
a flat-histogram kinetic growth algorithm, we have grown $10^9$ walks at length 512.
As a consequence of using a flat-histogram method with respect to the atmosphere,
we could boost the occurrence of zero-atmosphere walks roughly by a factor of three,
with minimal computational overhead.

\begin{figure}[ht!]
\centering
\includegraphics[width=0.9\textwidth]{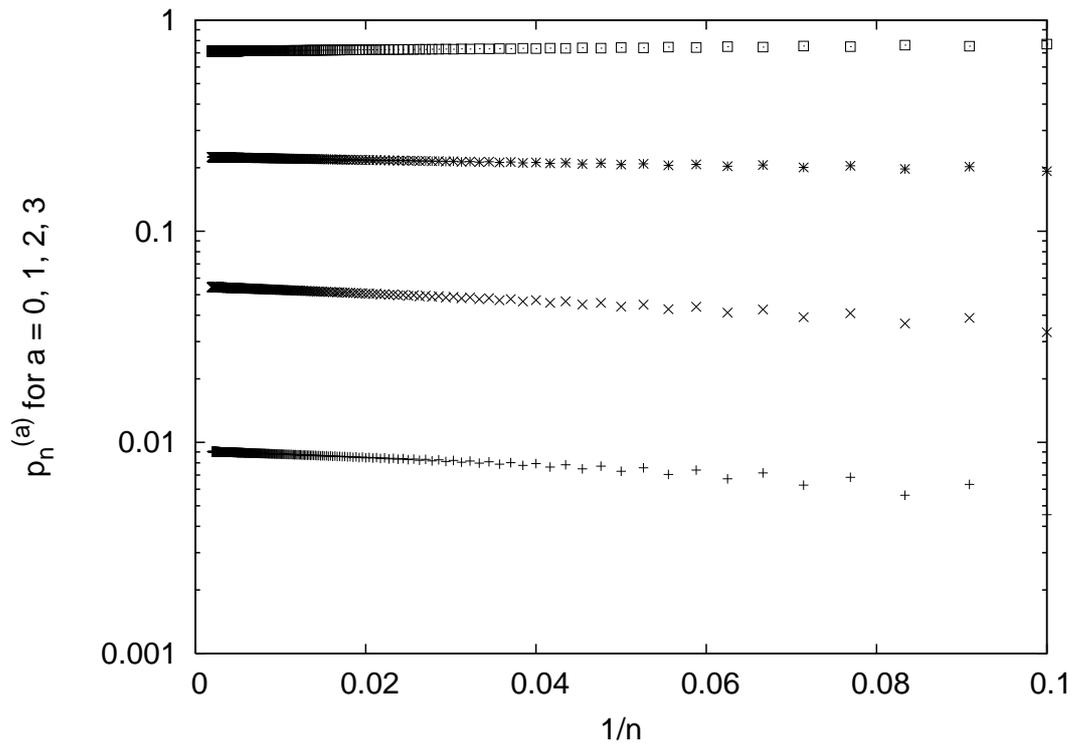}
\caption{A plot of the probability $p_n^{(a)}$ of an $n$-step self-avoiding walk on the
square lattice having atmosphere $a$. Shown are $p_n^{(a)}$ for $a=0$, $1$, $2$, $3$ 
(from bottom to top) on a logarithmic scale versus $1/n$. We note that there are only small
corrections to scaling.}
\label{figure2}
\end{figure}

\begin{figure}[ht!]

\centering
\includegraphics[width=0.9\textwidth]{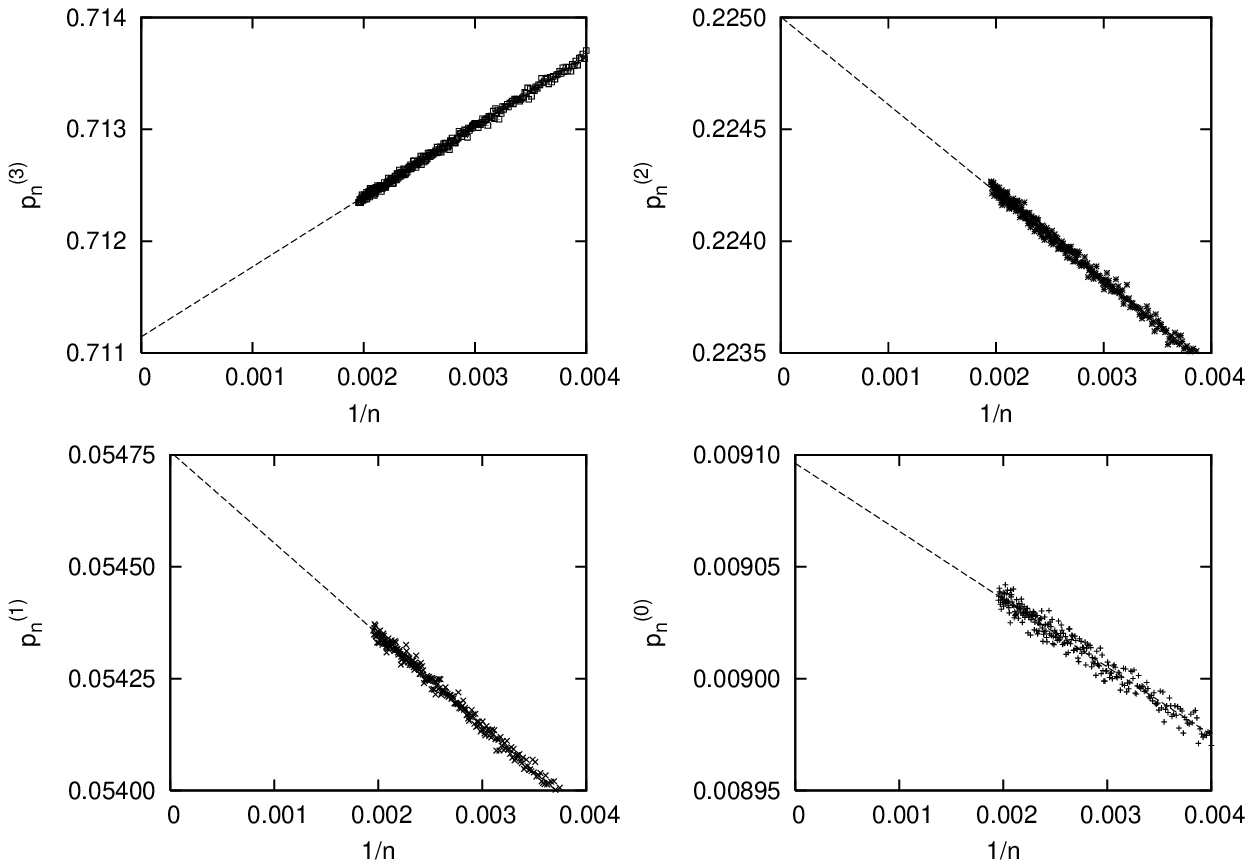}
\caption{A plot of the probability $p_n^{(a)}$ of an $n$-step self-avoiding walk on the
square lattice having atmosphere $a$. Shown are $p_n^{(a)}$ for $a=0$ (bottom right), $1$ (bottom left), $2$ (top right), $3$ (top left) on
a linear scale versus $1/n$, together with linear fits of $p_n^{(a)}$ to $1/n$ computed 
from all walks with at least $100$ steps.}
\label{figure3}
\end{figure}

From Figure \ref{figure2} it is clear that the quantities $p_{n}^{(a)}$ approach an
asymptotic value quickly as $n\to\infty$ with only small corrections to scaling.

Figure \ref{figure3} shows that these corrections are asymptotically linear in $1/n$.
We therefore conjecture on the basis of our simulations that the limit
$$p^{(a)}=\lim_{n\to\infty}\frac{c_n^{(a)}}{c_n}$$
indeed exists. A linear fit can be used to obtain estimates for $p^{(a)}$, and we find that
\begin{align}
p^{(0)}=&0.009096(4)\\
p^{(1)}=&0.05476(1)\\
p^{(2)}=&0.22500(2)\\
p^{(3)}=&0.71114(3)
\end{align}
for self-avoiding walks on the square lattice $\mathbb Z^2$.

\section*{Acknowledgements}
Financial support from the Australian Research Council via its support
for the Centre of Excellence for Mathematics and Statistics of Complex
Systems is gratefully acknowledged by the authors.


\end{document}